\documentclass[aps, pre, preprint, amsmath, amssymb, showpacs, superscriptaddress, 12pt, floats]{revtex4-1}

\usepackage[utf8]{inputenc}
\usepackage[russian, english]{babel}

\usepackage{graphicx}
\usepackage{mathtools}
\usepackage{textcomp}

\allowdisplaybreaks

\begin{document}

\title{Desynchronization in heterogeneous networks due to uniform exposure}

\author{V. A. Kostin}
%\email{vk1@ipfran.ru}
\affiliation{University of Nizhny Novgorod, Nizhny Novgorod 603950, Russia}
\affiliation{Institute of Applied Physics, Russian Academy of Sciences, Nizhny Novgorod 603950, Russia}
\author{G. V. Osipov}
\affiliation{University of Nizhny Novgorod, Nizhny Novgorod 603950, Russia}

\date{\today}
\begin{abstract}
  We study synchronization and rhythmic patterns generated in the heterogeneous cluster of FitzHugh–Nagumo oscillators with transition between self-oscillating and excitable elements.
  Such cluster models the sinoatrial node of the heart, and the particular point of focus is the role of different physiological responses to mechanical stress: stretch-activated current and conductivity change.
  The comparison between them finds the difference which can serve to identify the dominant mechanism in specific experimental situations.
\end{abstract}

\maketitle

The network of mixed self-oscillating and excitable elements arise in various areas of biology and physics~\cite{sinha_patterns_2014, hens_bursting_2015}.
Important examples of such networks are related to the cardiac pacemakers (such as a sinoatrial node) incorporated in the surrounding excitable cardiac muscle as well as cardiac cell cultures studied in experiments~\cite{kanakov_cluster_2007, kryukov_synchronization_2008, duverger_multicellular_2014, qu_nonlinear_2014, alonso_nonlinear_2016}.
Among the actual problems associated with such pacemaker nodes and cell cultures, there are possible effects of external mechanical action and underlying physiological mechanisms~\cite{kharkovskaia_2018}.
The two main mechanisms often considered involve the stretch-activated ionic channels and change of the local conductivity due to the tissue deformations~\cite{nash_electromechanical_2004, keldermann_modeling_2009, trayanova_cardiac_2011, weise_emergence_2012, kostin_transient_2016}.
In regard of the network models, these two mechanisms correspond to additional external current terms and to the change in the local coupling strength.

In this paper, we study the synchronization and rhythmic patterns in the mixed lattices of the self-oscillating and excitable FitzHugh–Nagumo elements as depending on the uniform local coupling strength and external current.
Two types of two-dimensional rectangular lattices are considered and compared: ones with the smoothly inhomogeneous transition between self-oscillating and excitable elements and ones with a cluster of the random self-oscillating FitzHugh–Nagumo elements.
The former may be associated with the sinoatrial node inside an excitable tissue, and the latter may describe an artificial culture of cardiac cells.

The studies are based on solving ordinary differential equations describing a two-dimensional rectangular lattice of the Fitzhugh–Nagumo elements numerically.
The corresponding equations are
\begin{gather*}
\dot V_{ij} = d \left(V_{i, j + 1} + V_{i, j - 1} + V_{i + 1, j} + V_{i - 1, j} - V_{ij}\right) - V_{ij} \left(V_{ij} - a_{ij}\right) \left(V_{ij} - 1\right) - W_{ij} + I,\\
\dot W_{ij} = \varepsilon \left(kV_{ij} - W_{ij}\right),
\end{gather*}
where $V_{ij}(t)$ and $W_{ij}(t)$ are the variables of the FitzHugh–Nagumo element with indices $i$ and $j$; $a_{ij}$, $k$, and $\varepsilon$ are the parameters of the FitzHugh–Nagumo model, all being homogeneous except the parameters $a_{ij}$; $I$ is the parameter describing uniform external current or current of the stretch-activated channels; $d$ is the local (diffusive) coupling strength; the indices $i$ and $j$ range from $0$ to $M - 1$ and $N - 1$.
In the equations for boundary elements, it is assumed that $V_{-1j} \equiv V_{0j}$, $V_{Mj} \equiv V_{M - 1, j}$, $U_{i, -1} \equiv V_{i0}$, and $U_{iN} \equiv V_{i, N - 1}$, as it would be for the Neumann boundary conditions.
Below, we consider square lattices with $N = M = 50$.
In simulations, the parameter values $k = 0.5$ and $\varepsilon = 0.01$ are chosen.
Two types of initial conditions were used: the uniform conditions $V_{ij}(0) = 0.1$, $U_{ij}(0) = 0$ and the random conditions $U_{ij}(0), V_{ij}(0) \sim \mathcal U(0, 1)$.
The used calculation time interval $T = 2000$ is large compared to the characteristic interval between spikes so at least several spikes are observed in most lattice elements.
For the element in the lattice corner (where the elements are excitable), the rhythmic pattern of $V_{M - 1, N - 1}(t)$ is analyzed within the time period $500 < t < 2000$, and the mean and standard deviation of the intervals between spikes are calculated as well as the variability of the rate equal to the ratio between the standard deviation and the average (the relative standard deviation).

First, let us consider the smoothly inhomogeneous lattice simulating a sinoatrial node with the surrounding muscle.
Around the corner $i = j = 0$, the elements are self-oscillatory, and in the opposite corner the elements are excitable.
The parameter $a_{ij}$ depends smoothly on the distance $\sqrt{i^2 + j^2}$ to the corner; it increases with $\sqrt{i^2 + j^2}$ from some negative value $a_{\min}$ to a positive one $a_{\max}$ at some distance $r_{\max}$, for distances greater than $r_{\max}$ the parameter $a_{ij}$ is uniform,
\begin{equation*}
  a_{ij} = \begin{cases}
    a_{\max} + \left(a_{\min} - a_{\max}\right) \cos^2\left(\frac{\pi \sqrt{i^2 + j^2}}{2r_0}\right), & \sqrt{i^2 + j^2} \leqslant r_0, \\
    a_{\max}, & \sqrt{i^2 + j^2} > r_0;
  \end{cases}
\end{equation*}
Figure~1 show the dependence $a_{ij}$.
\begin{figure}[h]
  \includegraphics[width=0.7\textwidth]{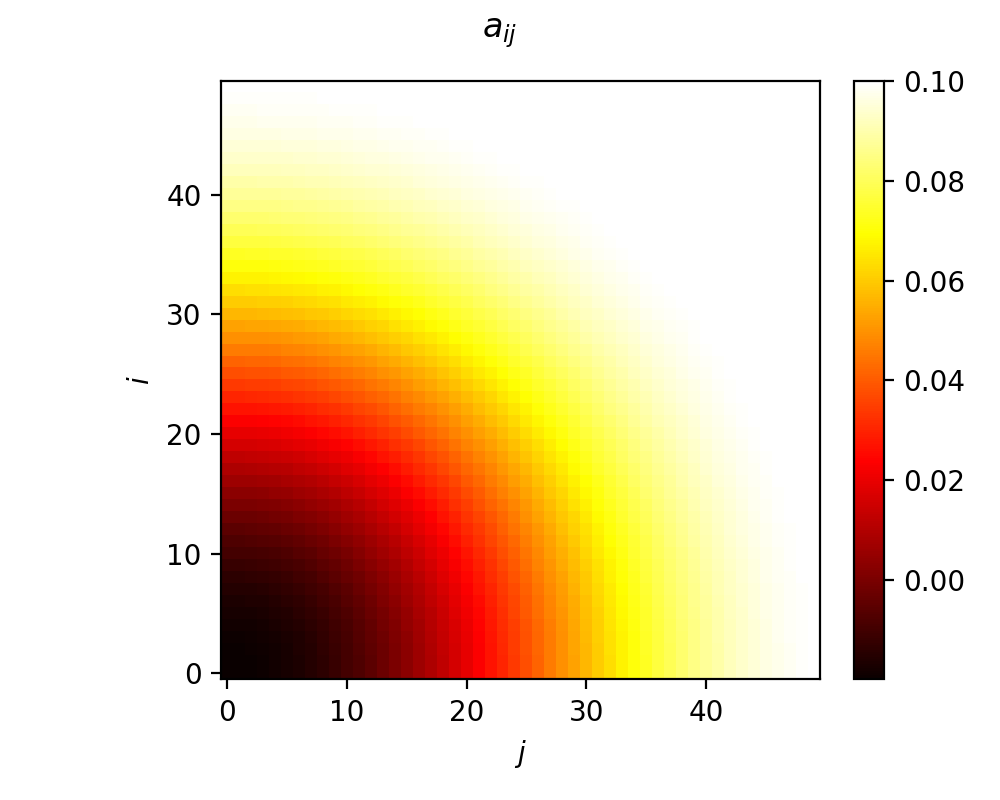}
  \caption{The smoothly inhomogeneous spatial profile of the parameter $a_{ij}$ with $a_{\min} = -0.1$, $a_{\max} = 0.02$, $r_0 = 50$.}
\end{figure}
In Fig.~2, one can see the calculation result for the lattice with the spatial profile of $a_{ij}$ shown in Fig.~1. in the case of the uniform initial conditions.
\begin{figure}[h]
  \includegraphics[width=1.0\textwidth]{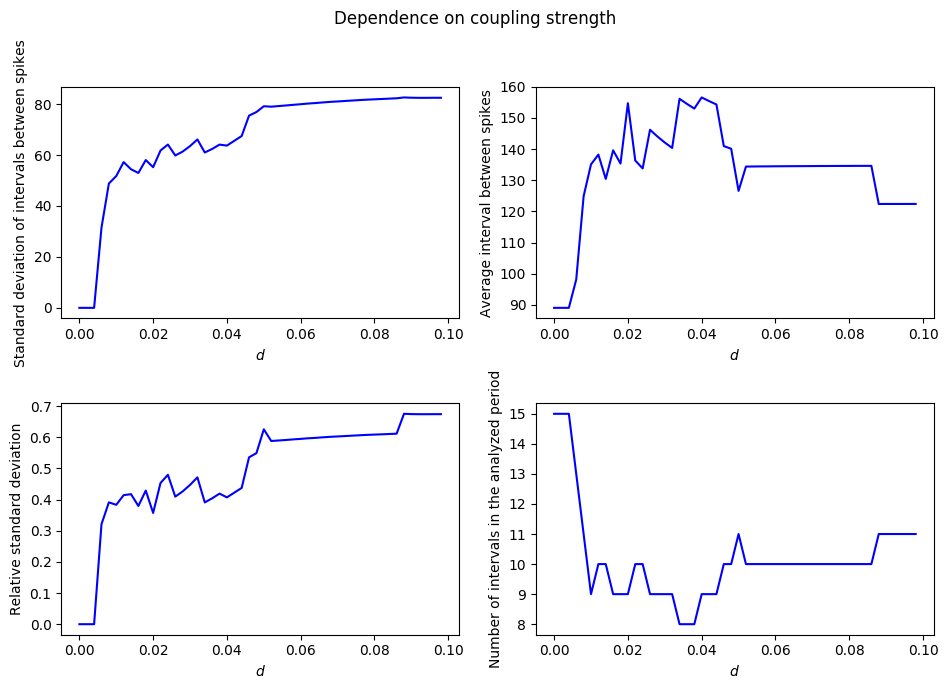}
  \caption{The characteristics of generated rhythmic pattern against the local coupling strength: standard deviation of the intervals between spikes, the average of these interval, the rhythm variability as the ratio between standard deviation and the average, and the number of spikes in the time period $500 < t < 2000$. 
           The initial conditions are uniform; the spatial profile of $a_{ij}$ is shown in Fig.~1; the external current is absent $I = 0$.}
\end{figure}
Figure~2 shows the dependences on the local coupling strength for the case  the characteristics of the rhythmic pattern generated in the top right corner by the oscillating elements in the bottom left corner.
One can see the trend of the variability decreasing with the coupling strength decreasing.
There are some stepwise jumps in the shown dependences indicating partial desynchronization, however the oscillating remains mainly synchronized until the coupling strength is so small that there is no excitation in the top right corner.

The situation differs drastically in the case of random initial conditions as shown in Fig.~3.
\begin{figure}[h]
  \includegraphics[width=1.0\textwidth]{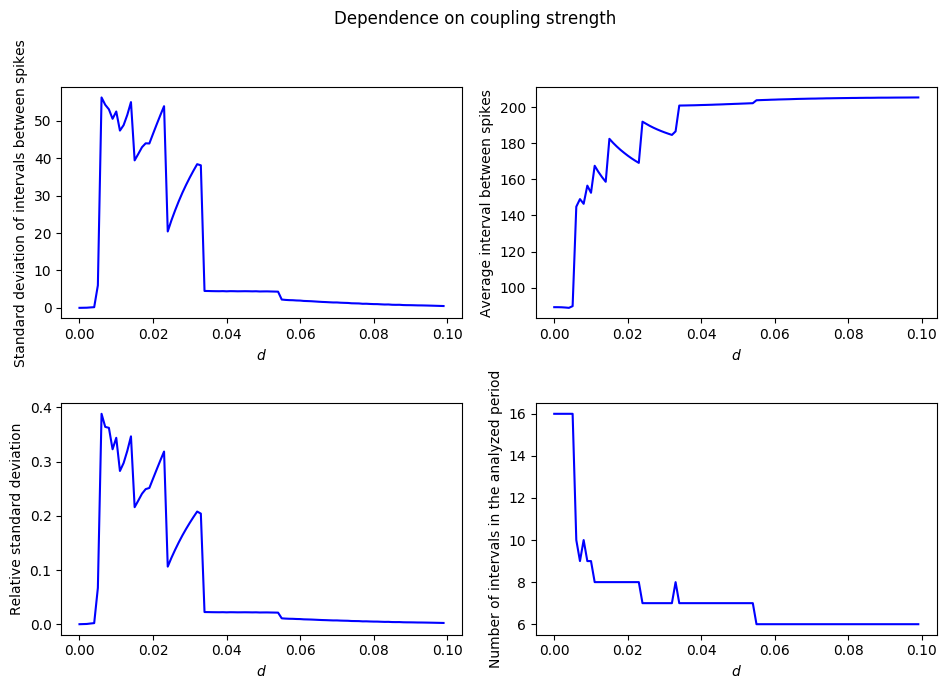}
  \caption{The same as in Fig.~2, but for the random initial conditions.}
\end{figure}
Here, there is a strong increase in the variability after the coupling strength goes under some critical value.
The multiple steps in the shown dependence can be attributed to the not too large computation time.
The average interval has a clear trend to decrease with decreasing coupling, which also differs from the previous case.
The difference between the case of uniform and random initial conditions indicates the likely multistability.
The lack of the synchronization at low coupling strength may be seen from Fig. 4, where the temporal dependences $V_{ij}(t)$ are shown for different elements in the lattice. 
\begin{figure}[h]
  \includegraphics[width=0.7\textwidth]{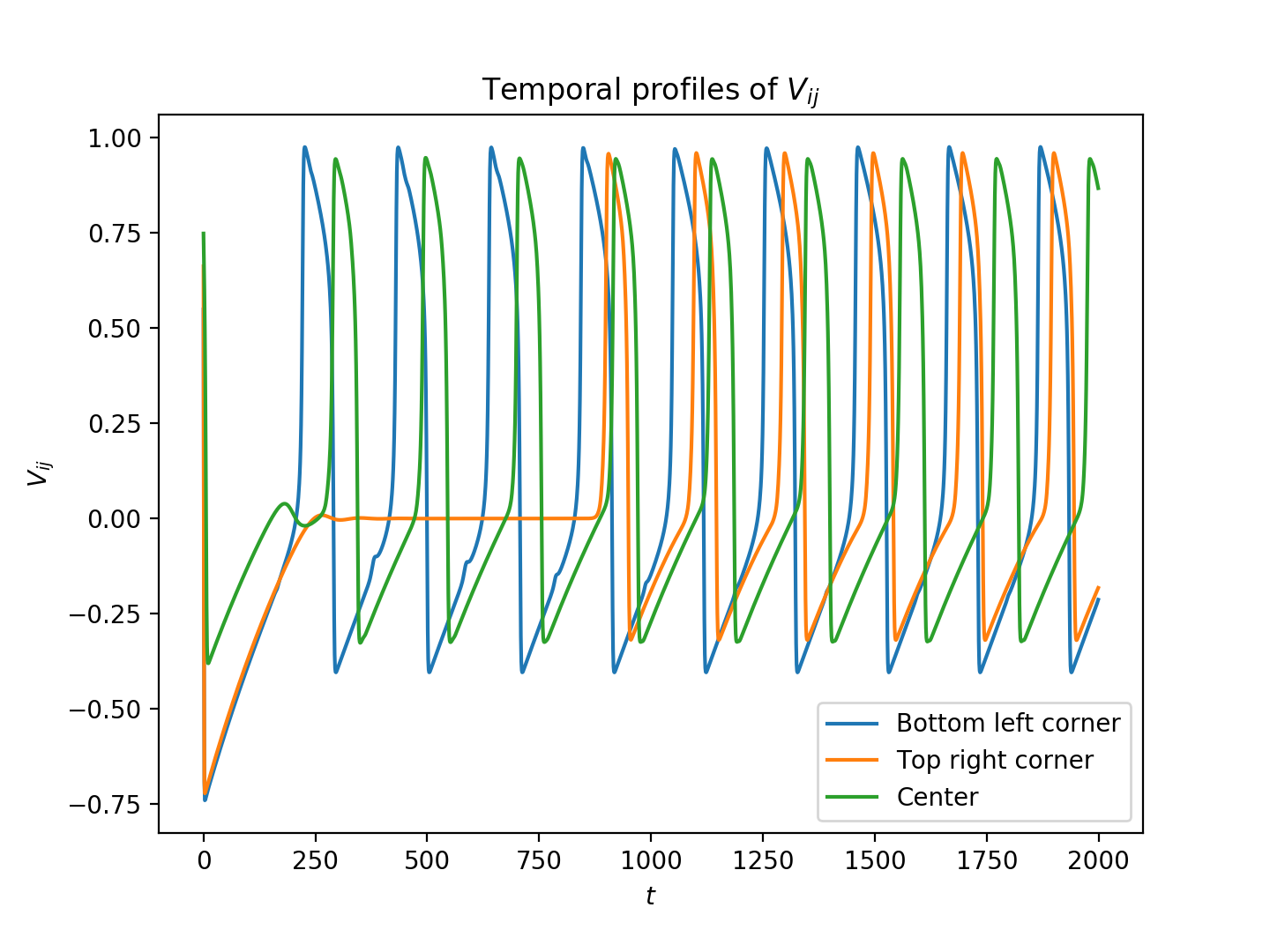}
  \caption{The temporal profiles $V_{00}(t)$ (the bottom left corner), $V_{25,25}(t)$ (the center), $V_{49,49}$ (the top right corner).
   The coupling strength is $d = 0.01$; the other parameters are as in Fig.~3.}
\end{figure}
The corresponding snapshot of the wave pattern is shown in Fig.~5.
\begin{figure}[h]
  \includegraphics[width=0.7\textwidth]{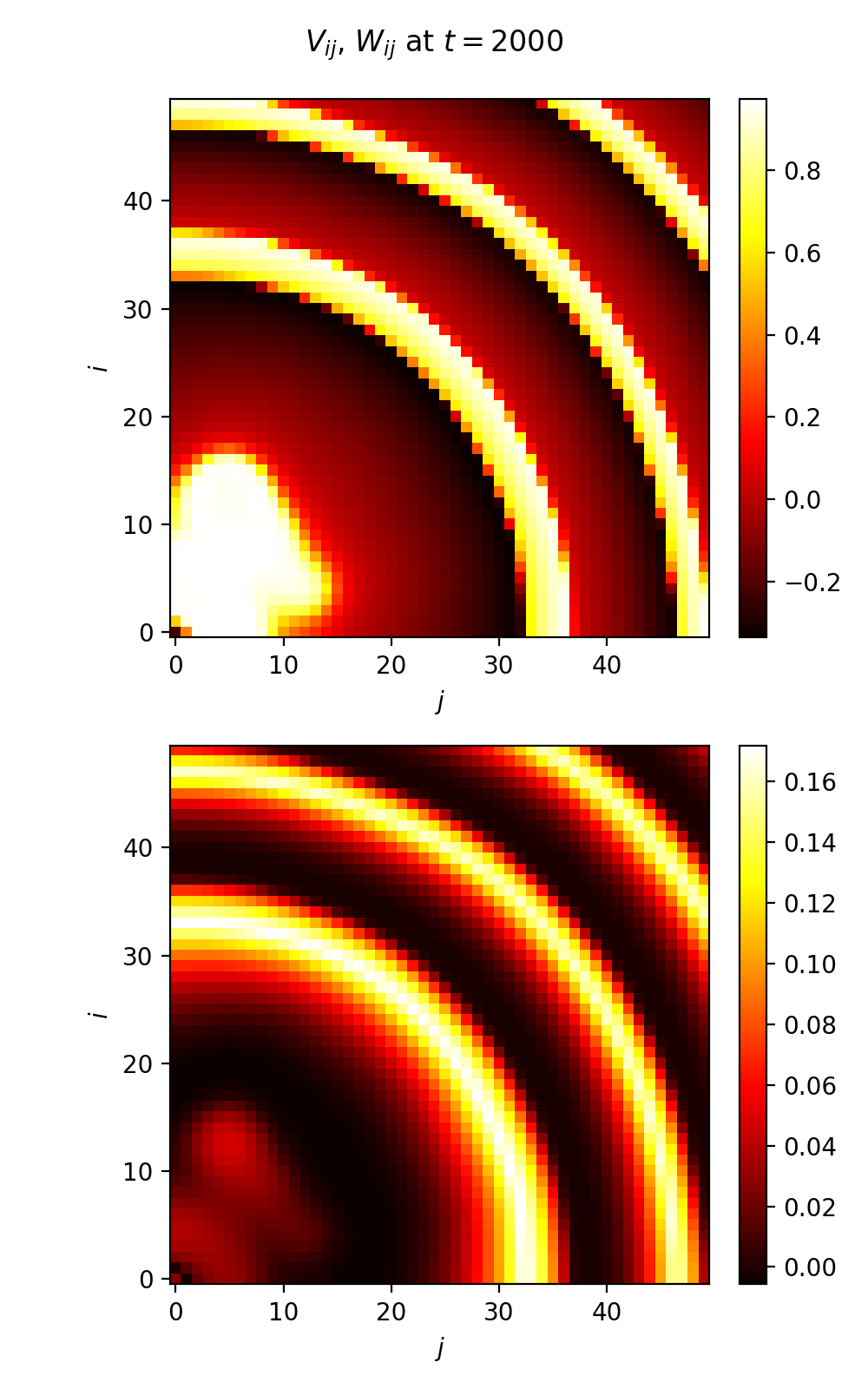}
  \caption{The snapshot of the wave pattern for $V_{ij}$ and $W_{ij}$ at $t = 2000$.
           The parameters are as in Fig.~4.}
\end{figure}

For the lattices with random cluster of oscillating elements, the dependences on the coupling strength are similar to the case with smoothly inhomogeneous lattices and random initial conditions.
In this case, the type of initial conditions affects the dynamics less than in the case with smoothly inhomogeneous lattice.
We choose the lattice with $a_{ij} = a_{\max}$ at $\sqrt{i^2 + j^2} \geqslant r_0$ and $a_{ij} \sim \mathcal U(a_{\min}, 0)$ at $\sqrt{i^2 + j^2} < r_0$ as shown in Fig.~6.
\begin{figure}[h]
  \includegraphics[width=0.7\textwidth]{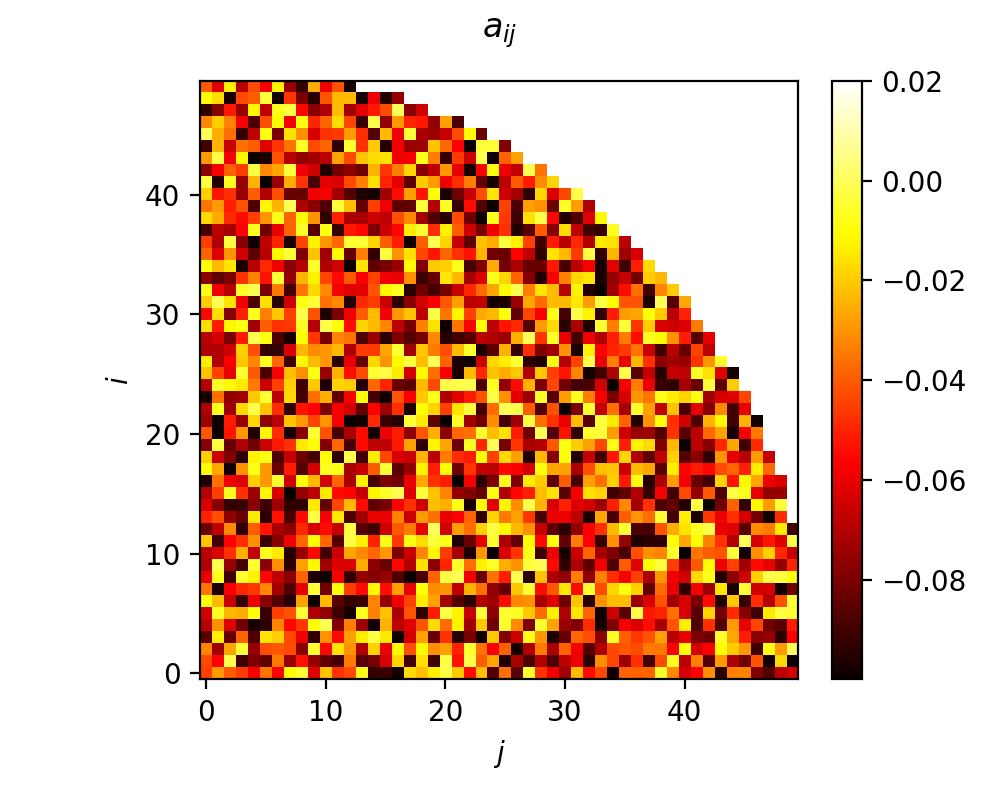}
  \caption{The spatial profile of the parameter $a_{ij}$ with the random cluster of self-oscillating elements, $a_{\min} = -0.1$, $a_{\max} = 0.02$, $r_0 = 50$.}
\end{figure}
In Fig.~7 one can see the calculation results with strong prominent increase in variability at low coupling strengths.
\begin{figure}[h]
  \includegraphics[width=1.0\textwidth]{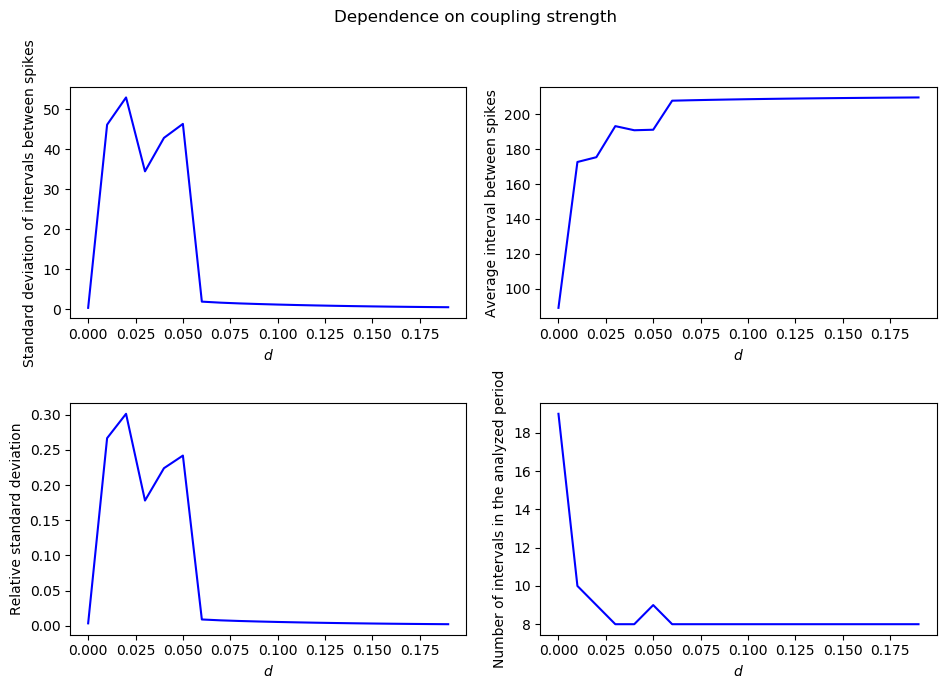}
  \caption{The same as in Fig.~3 but for the spatial profile of $a_{ij}$ shown in Fig.~6.}
\end{figure}

For the smoothly inhomogeneous lattices, the dependences of the rhythmic pattern on the external current was also calculated.
The dependences on the external current are more complicated and have probably some footprint from the specific model of the cardiac cells.
These dependences are highly non-monotonic and have multiple intervals (windows) of desynchronization, some narrow enough with prominent peaks in the variability as shown in Fig. 8.
\begin{figure}[h]
  \includegraphics[width=1.0\textwidth]{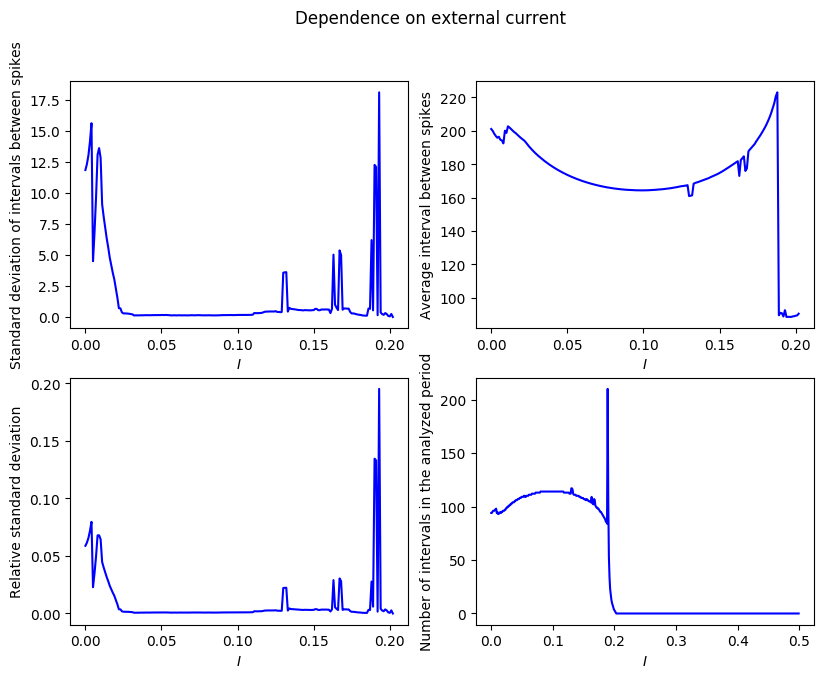}
  \caption{The characteristics of generated rhythmic pattern against the uniform external current: standard deviation of the intervals between spikes, the average of these interval, the rhythm variability as the ratio between standard deviation and the average, and the number of spikes in the time period $500 < t < 2000$. 
           The initial conditions are random; the spatial profile of $a_{ij}$ is shown in Fig.~1; the coupling strength is $d = 0.01$.}
\end{figure}
These windows are inherent to the collective dynamics of the ensemble and are not prominent in the individual dynamics.

To conclude, the desynchronization in inhomogeneous two-dimensional mixed-type lattices with excitable and self-oscillating FitzHugh–Nagumo elements is studied numerically depending on the type of lattice inhomogeneity (smooth or random), coupling forces and external current. 
The characteristics of the rhythm pattern (rate and its variability) created by clusters of oscillating elements in an excitable environment were calculated.
The uniformity of the initial state was shown to play an important role in desynchronization effects associated with weakening coupling.
The dependences of the rate and variability on the external current has complex structure with multiple narrow desynchronization regions.

\begin{acknowledgments}
  The work was supported by the Russian Science Foundation (Grant No. 14-12-00811).
\end{acknowledgments}

%\bibliography{elmechsyn}
%

\end{document}